\documentclass[useAMS,usenatbib]{mn2e}
\usepackage{epsfig}
\usepackage{amssymb}
\usepackage{amsmath}
\usepackage{graphicx}
\usepackage{caption}
\usepackage{tabto}
\usepackage{tabularx}
\usepackage{longtable}
\usepackage{lscape}
\usepackage{booktabs}
\usepackage{titlesec}
\usepackage{float}
\usepackage{url}
\usepackage{color}
\usepackage{subfigure}
\def\Msun{\ifmmode{{\rm M}_\odot}\else${\rm M}_\odot$\fi}
\def\msun{\ifmmode{{\rm M}_\odot}\else${\rm M}_\odot$\fi}
\def\Rsun{\ifmmode{{\rm R}_\odot}\else${\rm R}_\odot$\fi}
\def\rsun{\ifmmode{{\rm R}_\odot}\else${\rm R}_\odot$\fi}

\newcommand{\lsimeq}{\mbox{$\, \stackrel{\scriptstyle <}{\scriptstyle\sim}\,$}}
\newcommand{\gsimeq}{\mbox{$\, \stackrel{\scriptstyle >}{\scriptstyle\sim}\,$}}

\title[Origin of magnetic fields]
{Origin of magnetic fields in cataclysmic variables}

\author[Briggs, Ferrario, Tout, \& Wickramasinghe]
{Gordon P.~Briggs$^1$, Lilia Ferrario$^1$, Christopher A.~Tout$^{1,2,3}$,
\newauthor Dayal T.~Wickramasinghe$^1$\\
$^1$Mathematical Sciences Institute, The Australian National
University, ACT 0200, Australia\\
$^2$Institute of Astronomy, The Observatories, Madingley Road, Cambridge CB3 0HA\\
$^3$Monash Centre for Astrophysics, School of Physics and Astronomy, 10 College Walk, Monash University 3800, Australia
}

\begin{document}

\date{Accepted.  Received ; in original form}

\maketitle

\label{firstpage}

\begin{abstract}
  In a series of recent papers it has been proposed that high field
  magnetic white dwarfs are the result of close binary interaction and
  merging. Population synthesis calculations have shown that the
  origin of isolated highly magnetic white dwarfs is consistent with
  the stellar merging hypothesis. In this picture the observed fields
  are caused by an $\alpha-\Omega$ dynamo driven by differential
  rotation. The strongest fields arise when the differential rotation
  equals the critical break up velocity and result from the merging of
  two stars (one of which has a degenerate core) during common
  envelope evolution or from the merging of two white dwarfs. We now
  synthesise a population of binary systems to investigate the
  hypothesis that the magnetic fields in the magnetic cataclysmic
  variables also originate during stellar interaction in the common
  envelope phase. Those systems that emerge from common envelope more
  tightly bound form the cataclysmic variables with the strongest
  magnetic fields. We vary the common envelope efficiency parameter
  and compare the results of our population syntheses with
  observations of magnetic cataclysmic variables.  We find that common
  envelope interaction can explain the observed characteristics of
  these magnetic systems if the envelope ejection efficiency is
    low.

\end{abstract}


\section{Introduction}
\label{Introd}

Cataclysmic variables (CVs) are close interacting binaries generally
consisting of a low-mass main-sequence (MS) star transferring matter
to the white dwarf (WD) primary via Roche lobe overflow
\citep{Warner1995}. In non-magnetic or weakly magnetic systems, which
make up the majority of observed CVs, the hydrogen-rich material
leaving the secondary star from the inner Lagrangian point forms an
accretion disc around the white dwarf. It is estimated that some
$20-25$\,per cent of all CVs host a magnetic white dwarf
\citep[MWDs,][]{WickFer2000,fer2015a}.  These systems are the magnetic
cataclysmic variables (MCVs). Among MCVs we have the strongly magnetic
AM Herculis variables or polars. In polars the high magnetic field of
the white dwarf can thread and channel the material from the secondary
star directly from the ballistic stream to form magnetically confined
accretion funnels, so preventing the formation of an accretion
disc. In these systems the two stars are locked in synchronous
rotation at the orbital period. The radiation from the accretion
funnels \citep[e.g.][]{Ferrario1999} and the cyclotron radiation from
the shocks located at the funnels' footpoints of closed magnetic field
lines dominate the emission of these systems from the X-rays to the
infrared bands \citep[e.g.][]{MeggittWick1982,WickFer1988}.  Cyclotron
and Zeeman spectroscopy and spectropolarimetry have revealed the
presence of strong fields in the range of a few $10^7-10^8$\,G
\citep[see, e.g.,][]{Ferrario1992,Ferrario1993,Ferrario1996}.  Weaker fields of about $10^6$ to
$3\times 10^7$\,G are found in the DQ\,Herculis variables or
Intermediate Polars (IPs) where the white dwarf's magnetic field
cannot totally prevent the formation of an accretion disc
\citep[e.g. see][]{FWK1993}. In these systems the material is
magnetically threaded from the inner regions of a truncated accretion
disc and channelled on to the primary forming magnetically confined
accretion curtains \citep{FerrarioWick1993}. In the IPs the white
dwarf is not synchronously locked with the orbital period but is spun
up to a spin period shorter than the orbital period of the system.

\citet{Liebert2005} noticed the enigmatic lack of MWDs from the 501
detached binaries consisting of a white dwarf with a non-degenerate
companion found in the DR1 of the Sloan Digital Sky Survey
\citep[SDSS,][]{York2000}. They also noticed that among the 169 MWDs
known at the time, none had a non-degenerate detached companion. This
was puzzling because such a pairing is very common among non-magnetic
white dwarfs \citep[see, e.g.][]{Hurley2002,Ferrario2012}. A similar study
conducted on the much larger DR7 sample of SDSS detached binaries
consisting of a white dwarf with a non-degenerate companion
\citep{Kleinman2013} led to the same conclusion
\citep{Liebert2015}. Over the years, not a single survey conducted to
ascertain the incidence of magnetism among white dwarfs has yielded a
system consisting of a magnetic white dwarf with a non-degenerate
companion \citep[e.g.,][]{Schmidt2001,Kawka2007}. It is this curious
lack of pairing that led \citet{Tout2008} to propose that the
existence of magnetic fields in white dwarfs is intimately connected
to the duplicity of their progenitors and that they are the result of
stellar interaction during common envelope evolution.  In this
picture, as the cores of the two stars approach each other, their
orbital period decreases and the differential rotation that takes
place in the convective common envelope generates a dynamo mechanism
driven by various instabilities.  \citet{Regos1995} argued that it is
this dynamo mechanism that is responsible for the transfer of energy
and angular momentum from the orbit to the envelope which is
eventually, all or in part, ejected.

\citet{WTF2014} have shown that strong magnetic fields in white dwarfs
can be generated through an $\alpha-\Omega$ dynamo during common
envelope evolution where a weak seed poloidal field is wound up by
differential rotation to create a strong toroidal field. However
toroidal and poloidal fields are unstable on their own
\citep{Braithwaite2009}. Once the toroidal field reaches its maximum
strength and differential rotation subsides the decay of toroidal
field leads to the generation of a poloidal field with the two
components stabilising each other and limiting field growth until they
reach a final stable configuration. Thus, a poloidal seed field can be
magnified during common envelope evolution by an amount that depends
on the initial differential rotation but is independent of its initial
strength. According to this scenario the closer the cores of the two
stars are dragged at the end of common envelope evolution, before the
envelope is ejected, the greater the differential rotation and thus
the stronger the expected frozen-in magnetic field. If common envelope
evolution leads to the merging of the cores the result is an isolated highly
magnetic white dwarf. If the two stars do not coalesce they
emerge from the common envelope as a close binary that evolves into a
MCV.  The viability of such model, in terms of incidence of magnetism
among single white dwarfs and their mass and magnetic field
distribution, have been shown by \citet{Briggs2015} and
\citet{Briggs2018}, henceforth referred to as paper\,I and paper\,II
respectively.

In this paper we continue our studies of the origin of fields in MWDs
to explain the parentage of MCVs. To this end we carry out a
comprehensive population synthesis study of binaries for different
common envelope efficiencies $\alpha$. We examine all paths that lead
to a system consisting of a white dwarf with a low-mass companion
star.  We show that the observed properties of the MCVs are generally
consistent with their fields originating through common envelope
evolution for $\alpha<0.4$. 
 
\section{Evolution and space density of MCVs}

Observed MCVs consist of a white dwarf that accretes matter from a
secondary star that has not gone through any significant nuclear
evolution when the transfer of mass begins. The mass ratio of an MCV
is given by $q=M_{\rm sec}/M_{\rm WD}<1$ where $M_{\rm WD}$ is the
mass of the white dwarf primary and $M_{\rm sec}$ is the mass of the
companion star. The mass accretion process in MCVs is relatively
stable over long periods of time, although the polars suffer from high
and low states of accretion. It is not known what sparks the change in
accretion mode but, because polars do not have a reservoir of matter
in an accretion disc, they can switch very quickly from high to low
states. IPs have never been observed in low states of accretion. 
  Stable mass transfer can be driven by nuclear-timescale expansion of
  the secondary (not generally applicable to MCVs) and/or by loss of
  angular momentum, driven by magnetic braking above the CV period gap
  \citep[caused by the disrupted magnetic braking mechanism,
  see][]{Spruit1983, Rappaport1983, Verbunt1984} and gravitational
  radiation below the gap. Loss of angular momentum shrinks the orbit
  keeping the companion star filling its Roche lobe and so drives mass
  transfer. Therefore, as MCVs evolve, the orbital period diminishes
  until it reaches a minimum when the secondary star becomes a
  substellar-type object whose radius increases as further mass is
  lost. Systems that have reached the minimum period and have turned
back to evolve toward longer periods are often called period bouncers
\citep[e.g.][]{Patterson1998}

The evolution of MCVs is expected to be similar to that of
non-magnetic CVs. However, \citet{LiWuWick1994} have shown that
angular momentum loss may not be as efficient in polars as it is in
non-magnetic or weakly magnetic CVs in bringing the two stars together
because the wind from the secondary star is trapped within the
magnetosphere of the white dwarf. This phenomenon slows down
the loss of angular momentum, reduces the mass transfer rate and leads to
longer evolutionary timescales. This mechanism provides a simple
explanation for the observed high incidence of magnetic systems among
CVs \citep{Araujo2005}.
\begin{figure}
\includegraphics[width=1.05\columnwidth]{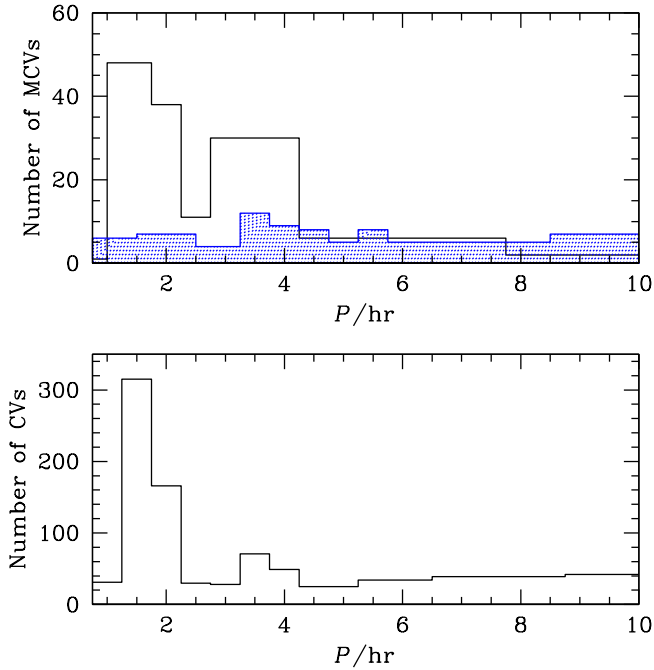}
\caption{The orbital period distribution of MCVs (top) and CVs
  (bottom). The MCVs are subdivided into Polars (solid black line histogram) and
  IPs (shaded histogram). We have used the latest version (v7.20) of
  Ritter \& Kolb's (2003) CV catalogue to create this figure.}
\label{fig:orbdistr}
\end{figure}
We show in Fig.\,\ref{fig:orbdistr} the period distribution of CVs and
MCVs where the MCVs have been subdivided into polars and intermediate
polars.  The space density of CVs is not well known and, over the
years, there has been some considerable disagreement between
observations and theoretical predictions. The recent study of
\emph{Swift}\,X-ray spectra of an optically selected sample of nearby
CVs conducted by \citet{Reis2013} has unveiled a number of very low
emission X-ray systems.  Hard X-ray surveys of the Galactic ridge have
shown that a substantial amount of diffuse emission can be
resolved into discrete low-luminosity sources.  Because the MCVs are
generally strong X-ray emitters, \citet{Muno2004} and \citet{Hong2012}
have propounded that IPs could be the main components of these
low-luminosity hard X-ray sources.

\citet{Pretorius2013} have conducted a study of the X-ray flux-limited
\emph{ROSAT} Bright Survey (RBS) to determine the space density of
MCVs. They assume that the 30 MCVs in the RBS are representative of
the intrinsic population. They also allow for a 50 per cent high-state
duty cycle for polars under the assumption that polars are below the
RBS detection threshold while they are in low states of
accretion. They find that the total space density of MCVs is
$1.3^{+0.6}_{−0.4}\times 10^{−6}$ pc$^{-3}$ with about one IP per
200\,000 stars in the solar neighbourhood. They conclude that IPs are
indeed a possible explanation for most of the X-ray sources in the
Galactic Centre. These new findings seem to suggest that the space
density of CVs is somewhat larger than initially forecast and thus in
closer agreement with theoretical expectations.

\subsection{Where are the progenitors of the MCVs?}\label{WhereProg}

\begin{figure}
\includegraphics[width=1.03\columnwidth]{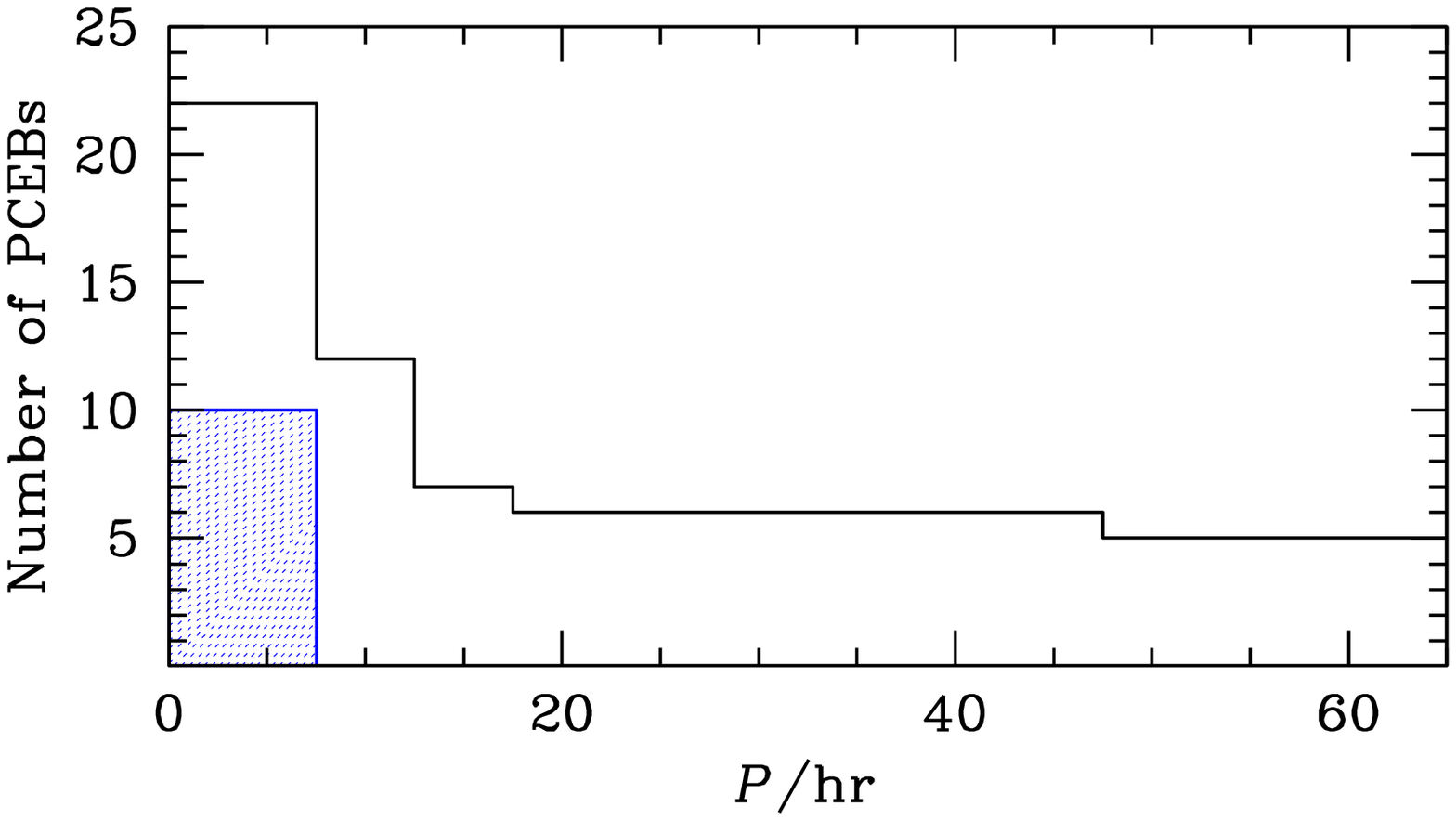}
\caption{The orbital period distribution of PCEBs
  \citep[solid black line histogram,][]{Nebot2011} and PREPs
  \citep[shaded histogram,][]{fer2015a}.}
\label{preps_fig}
\end{figure}

\citet{Liebert2005, Liebert2015} asked, ``Where are the magnetic white
dwarfs with detached, non-degenerate companions?''  This question is
awaiting an answer and thus the progenitors of the MCVs still need to
be identified.  The proposal by \citet{Tout2008} that the existence of
high magnetic fields in isolated and binary white dwarfs is related to
their duplicity prior to common envelope evolution is gaining
momentum. Observational support for the binary origin of magnetic
fields in MCVs is also strengthening. \citet{Zorotovic2010} listed
about $60$ post common envelope binaries (PCEBs) from the SDSS and
other surveys consisting of a white dwarf with an M-dwarf
companion. The periods of these PCEBs range from about $0.08$ to
$20$\,d and the white dwarf effective temperatures in the range 7\,500
to 60\,000\,K. According to current binary evolution theory, one third
of these systems should lose angular momentum from their orbits by
magnetic braking and gravitational radiation and are expected to come
into contact within a Hubble time. However none of these 60 binaries
contains a MWD, even if observations indicate that 20 to 25\,per cent
of all CVs harbour one. This finding suggests that magnetic white
dwarf primaries are only present in those binaries that are already
interacting or are close to interaction. The magnetic systems
originally known as Low-Accretion Polars
\citep[LARPS,][]{Schwope2002,FWS2005} have been proposed to be the
progenitors of the polars. The first LARPS were discovered in the
Hamburg/ESO Quasar Survey \citep[HQS,][]{Wisotzki1991} and then by the
SDSS by virtue of their unusual colours arising from the presence of
strong cyclotron harmonic features in the optical band together with a
red excess owing to the presence of a low-mass red companion star. The
MWDs in LARPS are generally quite cool
($T_{\rm eff}\lsimeq10\,000$\,K) and have low-mass MS companions which
underfill their Roche lobes \citep[e.g.][]{Reimers1999, Schwope2002,
  FWS2005, Parsons2013}. The MWDs in these systems accrete mass from
the wind of their companion at a rate substantially larger than in
detached non-magnetic PCEBs \citep{Parsons2013}. \citet{Tout2008}
suggested that LARPS could be pre-polars waiting for gravitational
radiation to bring the stars close enough to each other to allow Roche
lobe overflow to commence. These systems were renamed pre-polars
(PREPs) by \citet{Schwope2009} to avoid confusion with polars in a low
state of accretion. PREPs have orbital periods which are, on average,
only marginally longer than those of polars. The ages of the white
dwarfs in PREPs, as indicated by their effective temperatures, are
generally above a billion years. The absence of PREPs with hot white
dwarfs is puzzling but maybe still not alarming, if one considers the
small number of PREPs currently known and the initial rapid cooling of
white dwarfs. Thus, the hypothesis that the progenitors of MCVs are
expected to emerge from common envelope when they are close to
transferring mass via Roche Lobe overflow is well warranted. We show
in Fig.\,\ref{preps_fig} the period distribution of PCEBs and PREPs.

\section{Population synthesis calculations}\label{popsynt}
\label{sec:calculations}

\begin{table*}
  \caption{We have indicated with $N$ (second column) the fraction of PREPs for different efficiency parameters $\alpha$ (first column) in a single generation of binaries. The other columns give the smallest and the largest progenitor masses and initial orbital periods.}
  \centering
\begin{tabular}{ c c c c r c c r}
\hline
\noalign{\smallskip}
$\alpha$ & $N$\,(per cent) & $M_{1_{\rm min}}/\msun$ & $M_{2_{\rm
    min}}/\msun$ &$M_{1_{\rm max}}/\msun
$&$M_{2_{\rm max}}/\msun $ & $P_{0_{\rm min}}/$d & $P_{0_{\rm max}}/$d \\
\noalign{\smallskip}
\hline
\noalign{\smallskip}
0.10   & 1.518  & 1.08 & 0.10 & 8.16 & 1.42   & 369.7 &  3144.0  \\
0.15   & 1.672  & 1.08 & 0.10 & 8.16 & 1.42   & 293.3 &  2800.5  \\
0.20   & 1.663  & 1.08 & 0.10 & 8.16 & 1.42   & 246.6 &  2354.3  \\
0.25   & 1.213  & 1.08 & 0.10 & 8.16 & 1.36   & 207.3 &  2097.0  \\
0.30   & 1.163  & 1.08 & 0.10 & 8.16 & 1.14   & 184.6 &  2221.9  \\ 
0.50   & 0.808  & 1.08 & 0.10 & 8.16 & 0.58   & 123.2 &  2221.9  \\
0.70   & 0.804  & 1.08 & 0.10  & 8.16 & 0.19  &   87.0 &  1867.9  \\
0.80   & 0.859  & 1.08 & 0.10  & 8.16 & 0.13  &   69.1 &  1762.9  \\
\noalign{\smallskip}
\hline
\end{tabular}
\label{tab:statsCE}
\end{table*}

Each binary is assigned three initial parameters. These are the mass
$1.0\le M_1/\msun\le 10.0$ of the primary star, the mass
$0.1\le M_2/\msun\le 2.0$ of the secondary star, and the orbital
period $1\le P_0/{\rm d}\le 10\,000$ at the zero-age main sequence
(ZAMS). We set the eccentricity to zero. We sampled each parameter
uniformly on a logarithmic scale with 200\,divisions.  This sampling
gives a synthetic population of about 70 million binary systems.  The
actual number of binary systems is then calculated on the premise that
$M_1$ follows Salpeter's mass function distribution
\citep{Salpeter1955} and $M_2$ is according to a flat mass ratio
distribution with $q\le 1$.  The initial period distribution is
assumed to be uniform in the logarithm.

We have used the rapid binary star evolution algorithm, {\sc bse},
developed by \citet{Hurley2002}, to evolve each binary system from the
ZAMS to $9.5\,$Gyr \citep[age for the Galactic
Disc,][]{Kilic2017}. {\sc bse} is an extension of the single star
evolution code written by \citet{Hurley2000}.  It allows for stellar
mass loss and interaction between the two stars such as mass transfer, Roche
lobe overflow, common envelope evolution \citep{Paczynski1976}, tidal
interaction, supernova kicks, and angular momentum loss caused by
gravitational radiation and magnetic braking.

In our modelling we have used the $\alpha$ formalism for common envelope
  evolution. If $\Delta E_{\rm orb}$ is the change in orbital energy during
the in-spiral phase and $E_{\rm bind}$ is the energy required to eject the
envelope then
\[
\Delta E_{\rm orb}=\alpha E_{\rm bind}.
\] 
\noindent Here $\alpha$ is the common envelope efficiency parameter,
ranging between 0.1 and 0.9, that takes into consideration the fact
that the removal of the envelope is not totally efficient (see
papers\,I and II for more details). The expression for the envelope
binding energy contains another parameter $\lambda$ which denotes the
structure, or central condensation, of the envelope of the donor. This
parameter is very uncertain and could vary by up to an order of
magnitude \citep{TaurisDewi2001, Ivanova2011}. It depends on both the
envelope structure and on the mass above which the envelope is
expelled. In this work we follow \citet{Hurley2002} and set
$\lambda=0.5$.  A review on common envelope evolution and on the
parameters that are used in its modelling can be found in
\citet{Izzard2012}.

Single star mass loss rates are described by \citet{Hurley2000}. In
our calculations we have adopted $\eta=1.0$ for the Reimers' mass-loss
parameter, as outlined in paper\,I, and a stellar metallicity
$Z = 0.02$.

Our theoretical sample of PCEBs consists of systems that (i)\,have
undergone common envelope evolution, (ii)\,have a primary that evolves
into a white dwarf, (iii)\,have a companion that remains largely
unevolved and (iv)\, have a mass ratio $q\le 1$.  A subset of these
systems come into contact over the age of the Galactic Disc and become
classical CVs. Those systems with a white dwarf that develops a strong
magnetic field become MCVs.

Of our sample of PCEBs, we then select the subset consisting of the
PREPs (the progenitors of the MCVs). PREPs must fulfil two additional
criteria: (i) the primary star must have a degenerate core before
entering the last common envelope phase and (ii) no further core
burning occurs. The reason for the first criterion is that a
degenerate core is essential for a stellar magnetic field to persist,
in a frozen-in state, after its formation.  The reason for the second
is that nuclear burning in the core would ignite convection that would
destroy any frozen-in magnetic field. Systems that violate either
criterion but come into contact over the age of the Galactic Disc are
expected to evolve into classical non-magnetic CVs. We show in
Table\,\ref{tab:statsCE} the limits of the parameter space within
which PREPs are formed.  The minimum ZAMS masses of the systems that
give rise to PREPs are listed in the columns with headings
$M_{1_{\rm min}}$ and $M_{2_{\rm min}}$ and the maximum masses are
under the headings $M_{1_{\rm max}}$, $M_{2_{\rm max}}$.  Minimum and
maximum initial periods are in the columns under $P_{0_{\rm min}}$ and
$P_{0_{\rm max}}$ respectively.

Once we have obtained our theoretical PREP sample, we assign a
magnetic field $B$ to each of their white dwarf primaries following
the prescription described in paper\,II to model the field
distribution of high field magnetic white dwarfs (HFMWDs). That is
\begin{equation}\label{EqBfield}
B = B_0\left(\frac{\Omega}{\Omega_{\rm crit}}\right)\, \mbox{G}.
\end{equation}
where $\Omega$ is the orbital angular velocity and
$\Omega_{\rm crit}=\displaystyle{\sqrt{GM_{\rm WD}/R_{\rm WD}^3}}$ is
the break-up angular velocity of a white dwarf of mass $M_{\rm WD}$
and radius $R_{\rm WD}$.  The parameter $B_0$ is a free parameter that
was determined empirically in paper\,II, that is,
$B_0=1.35\times 10^{10}$\,G. The parameter $B_0$ does not influence
the shape of the field distribution which is only determined by
$\alpha$. Lower (or higher) $B_0$ shift the field distribution to
lower (or higher) field strengths. Unlike HFMWDs, both stars emerge
from common envelope evolution but on a much tighter orbit that 
allows them to come into contact within a Hubble time and appear as MCVs.
Thus, white dwarfs in interacting binaries can only attain a fraction
of the upper field strength of single white dwarfs and this is the
reason why $B_0$ must be determined through the modelling of HFMWDs
(see paper\,II). Field strengths of MCVs are scaled down from the
maximum by equation (\ref{EqBfield}). 

\section{Synthetic population statistics}\label{SyntPop}

We have time integrated each population, characterised by
$\alpha$, to the Galactic Disc age under the assumption that
the star formation rate is constant. We have listed in
Table\,\ref{tab:PCEBS_PREPS_Percent} the percentage by type of all
binaries that emerge from common envelope over the age of the Galactic
Disc.

\begin{table}
  \caption{The number of PCEBs born, the fraction of PREPs from PCEBs
    and of MCVs (magnetic systems already exchanging mass) from PREP as a
    function of the common envelope efficiency parameter $\alpha$ over the
    age of the Galactic Disc. The number of PREPs is maximum close to
    $\alpha = 0.15$ while the number of MCVs is maximum at $\alpha= 0.10$.}
\centering
\begin{tabular} {l l l l}
\hline
\noalign{\smallskip}
$\alpha$ & Number of & \underline{PREPs} $\times100$& \underline{MCV~~} $\times100$ \\
              & PCEBs~~~   &PCEBs~~         & PREPS~   \\
\noalign{\smallskip}
\hline
\noalign{\smallskip}
0.10 & 30517472 &20.9 & 61.0 \\ 
0.15 & 36099023 &18.9 & 56.4 \\
0.20 & 38666876 &15.3 & 49.9 \\
0.30 & 41197674 &  8.7 & 45.0 \\
0.40 & 43654871 &  5.6 & 48.0 \\
0.50 & 46289395 &  4.5 & 51.0 \\
0.60 & 49010809 &  4.1 & 52.0 \\
0.70 & 51888317 &  3.8 & 52.4 \\
0.80 & 54664759 &  3.3 & 52.4 \\
\hline
\end{tabular}
\label{tab:PCEBS_PREPS_Percent}
\end{table}

Column 2 in Table\,\ref{tab:PCEBS_PREPS_Percent} shows that while the
number of PCEBs increases when $\alpha$ increases, the percentage of
PREPs (progenitors of the MCVs) decreases. This is because as $\alpha$
increases the envelope's clearance efficiency increases causing the
two stars to emerge from common envelope at wider separations and thus
less likely to become PREPs and thence MCVs. On the other hand, the
overall number of PCEBs increases because stellar merging events
become rarer at high $\alpha$, as shown in paper\,I.  Fewer merging
events are also responsible for the high incidence of systems with 
low mass He\,white dwarfs (He\,WDs) whose ZAMS progenitors were born at short orbital
periods and entered common envelope evolution when the primary star
became a Hertzsprung gap or a red giant branch (RGB) star. At larger
initial orbital periods common envelope evolution may occur on the
asymptotic giant branch (AGB). However as $\alpha$ increases only
stars in those systems that harbour massive enough white dwarfs can
come sufficiently close to each other to allow stable mass transfer to
occur within the age of the Galactic Disc (see section
\ref{mass_distribution}). In contrast, at low $\alpha$ the
clearance efficiency is low and so there is a longer time for the
envelope to exert a drag force on the orbit. This results in (i)
more merging events, (ii) tighter final orbits for all white dwarf
masses and (iii) a larger number of systems coming into contact over
the age of the Galactic Disc. Point (i) reduces the overall number of
PCEBs while (ii) and (iii) increase the number of PREPs.

\begin{table*}
  \caption{Evolutionary history of an example binary system that
    becomes a MCV after common envelope evolution with $\alpha=0.1$.
    Here CE\,=\,Common Envelope, RLO\,=\,Roche Lobe Overflow. }
  \centering
\begin{tabular}{ r r r r r r c l  }
\hline
\noalign{\smallskip}
Stage  & Time/Myr & $M_1/\msun$ & $M_2/\msun$ & $P/$\,d & $a/{\rm R}_\odot$ &$B/{\rm G}$& Remarks \\
\noalign{\smallskip}
\hline 
\noalign{\smallskip}
  1    &       0.000    &  4.577    & 0.230  & 2244.627 & 1218.030    & 0.000E+00   & ZAMS  \\  
  2    &    128.515   &  4.577    & 0.230  & 2244.627 & 1218.030    & 0.000E+00   & S1 is a Hertzsprung gap star \\
  3    &    129.078   &  4.577    & 0.230  & 2245.210 & 1218.188    & 0.000E+00   & S1 is a RGB star.  Separation increases slightly. \\
  4    &    129.445   &  4.574    & 0.230  & 2247.427 & 1218.790    & 0.000E+00   & S1 starts core He burning. Some mass loss occurs.\\
  5    &    149.930   &  4.466    & 0.230  & 2352.896 & 1247.059    & 0.000E+00   & S1 is an AGB star. Further mass loss occurs. \\  
  6    &    150.947   &  4.390    & 0.230  & 2173.184 & 1176.321    & 0.000E+00   & S1 is a late AGB star. Separation decreases significantly \\
  7    &    150.989   &  4.364    & 0.230  &  861.296  & 633.510      & 0.000E+00   & RLO \& CE start. Separation decreases dramatically. \\
  8    &    150.989   &  0.918    & 0.230  &    0.117    &  1.053        & 1.218E+07   & S1 emerges from CE as a CO\,MWD and RLO ceases.\\
  9    &    326.073   &  0.918    & 0.230  &    0.099    &  0.945        & 1.218E+07   & Separation decreases and MCV phase starts \\
 10   &  9\,500.000 &  0.918    & 0.037  &    0.139    &  1.112        & 1.218E+07   & Separation reaches a minimum between stages 9 and 10 \\
        &                   &               &           &                &                  &                     & and increases again. S2 is a brown dwarf. \\
\hline
\end{tabular}
\label{tab:evolMCV1}
\end{table*}

\begin{table*}
  \caption{Evolutionary history of a second example binary system that
    becomes a MCV after common envelope with $\alpha=0.4$.}
\centering
\begin{tabular}{ r r r r r r c l }
\hline
\noalign{\smallskip}
Stage  & Time/Myr & $M_1/\msun$ & $M_2/\msun$ & $P$\,(d) & $a/{\rm R}_\odot$ &$B/{\rm G}$& Remarks \\
\noalign{\smallskip}
\hline
\noalign{\smallskip}
  1  &          0.000 & 1.612  &  0.257  &   190.661  &  171.774   &    0.000E+00  & ZAMS  \\ 
  2  &    2197.329 &  1.612  &  0.257  &   190.661  &  171.774   &   0.000E+00  & S1 is a Hertzsprung gap star \\ 
  3  &    2239.430 &  1.611  &  0.257  &   190.743  &  171.811   &   0.000E+00  & S1 is a RGB star, loses mass.  Separation increases slightly. \\
  4  &    2343.048 &  1.580  &  0.257  &   110.351  &  118.629   &   0.000E+00  & S1 loses more mass, separation decreases. \\
  5  &    2343.048 &  0.386  &  0.257  &     0.149    &     1.020   &   3.577E+07  & RLO \& CE start. Separation decreases dramatically. \\
  6  &    2343.048 &  0.386  &  0.257  &     0.149    &     1.020   &   3.577E+07  & S1 emerges from CE as a He\,MWD and RLO ceases.\\
  7  &    3389.278 &  0.386  &  0.257  &     0.102    &     0.792   &   3.577E+07  & Separation decreases and MCV phase starts \\
  8  &  9\,500.000 &  0.386  &  0.052  &     0.100    &     0.687   &   3.577E+07  & Separation reaches a minimum between stages 7 and 8 \\
      &                   &            &            &                  &                &                      & and increases again. S2 is a brown dwarf.\\
\noalign{\smallskip}
\hline
\end{tabular}
\label{tab:evolMCV2}
\end{table*}

\subsection{Magnetic CV evolution examples}

The evolutionary history of a binary system depends on the parameters
that characterise it. The number of common envelope events can vary from one to
several \citep{Hurley2002}.  Whether a classical CV becomes magnetic
or not depends on the evolution before and after the common
envelope. Here we give two typical examples of systems that evolve
into a MCV. In the first the initially rather massive primary
star evolves into a CO\,white dwarf (CO\,WD) after common envelope evolution as
a late AGB star. In the second example the primary evolves into a
He\,white dwarf after  common envelope evolution while ascending the RGB.

\emph{Example 1:} Table~\ref{tab:evolMCV1} illustrates the evolution
of a system that becomes a close binary after common envelope with
$\alpha=0.1$.  The progenitors are a primary star (S1) of
$4.58\,\msun$ and a secondary star (S2) of sub-solar mass
$0.230\,\msun$.  At ZAMS the initial period is $2\,240\,$d with a
separation of $1\,220\,R _\odot$.

S1 evolves off the ZAMS and reaches the early AGB stage at
$149$\,Myr having lost $0.111$\,$\msun$ on the way.  After a
further $1.02$\,Myr S1 has become a late AGB star. Further evolution
brings the stars closer together at a separation of
$634$\,R$_\odot$. Soon after dynamically unstable Roche lobe
overflow from S1 to S2 takes place and common envelope begins.  At the
end of the short period of common envelope evolution the two stars
emerge with a separation of only $1.05$\,R$_\odot$ because of the
large orbital angular momentum loss during this stage.  The ejection
of the envelope exposes the core of S1 that has now become a magnetic
$0.918$\,$\msun$ CO\,WD.  After a further $175$\,Myr the
separation has further contracted to $0.945$\,R$_\odot$ via magnetic
braking and gravitational radiation.  Roche lobe overflow begins and
the system becomes a bona fide mass-exchange MCV. During the MCV
evolutionary phase the mass of the donor star, separation and orbital
period steadily decrease until the mass of the companion star becomes
too low to maintain hydrogen burning and S2 becomes a degenerate
object. At this point separation and orbital period reach a
minimum. Further evolution sees these two quantities increase again
over time. At an age of $9\,500$\,Myr S2 has lost most of its mass and
has become a $0.037$\,$\msun$ brown dwarf with the separation from its white dwarf
primary increased to $0.112$\,R$_\odot$.

\emph{Example 2:} Table~\ref{tab:evolMCV2} shows the evolution of a
second system that becomes a close binary after common envelope. This
time we have $\alpha=0.4$.  The progenitors are a MS
primary star (S1) of $1.61$\,$\msun$ and a secondary star (S2) of mass
$0.257$\,$\msun$.  At ZAMS the initial period is $191$\,d and the
separation $172$\,R$_\odot$.

S1 evolves off the ZAMS through the Hertzsprung gap to reach the RGB
after $2\,240$\,Myr having lost $0.001$\,$\msun$ on the way. Still on
the RGB at $2\,340$\,Myr S1 has lost $0.031$\,$\msun$ and the
separation has decreased to $119$\,R$_\odot$. Roche lobe overflow
from S1 to S2 and common envelope evolution begin. S1 emerges from
common envelope as a magnetic He\,WD with a mass of
$0.386$\,$\msun$. The orbital separation has drastically decreased to
$1.02$\,R$_\odot$. S2 maintains its mass and remains an M-dwarf
star. From this time onwards magnetic braking and gravitational radiation
cause the orbit to shrink further until at $3\,390$\,Myr the 
separation is $0.792$\,R$_\odot$ and Roche lobe overflow
commences. The system is now a MCV.  Further evolution leads S2 to
lose mass, owing to accretion on to S1, until, at $9\,500$\,Myr, S2 has
become a brown dwarf of mass $0.052$\,$\msun$ and the separation is
$0.687$\,R$_\odot$.

\subsection{Property distributions of the synthetic population}

We create our population of putative PREPs by integration over time
from $t=0$ to $t=9.5\,$Gyr.  The star formation rate is taken to be
constant over the age of the Galactic Disc.  Whereas
Table\,\ref{tab:statsCE} shows the relative numbers of PREPs obtained
from a single generation of binaries, continuous star formation
over the age of the Galactic Disc builds up a population of PCEBs,
PREPs, CVs, and MCVs that, as birth time increases, favours systems with
progressively higher mass primaries because lower mass primaries,
especially in later generations, do not have enough time to evolve to
the white dwarf stage.

\begin{figure*}
\begin{center}
\includegraphics[width=1.01\textwidth]{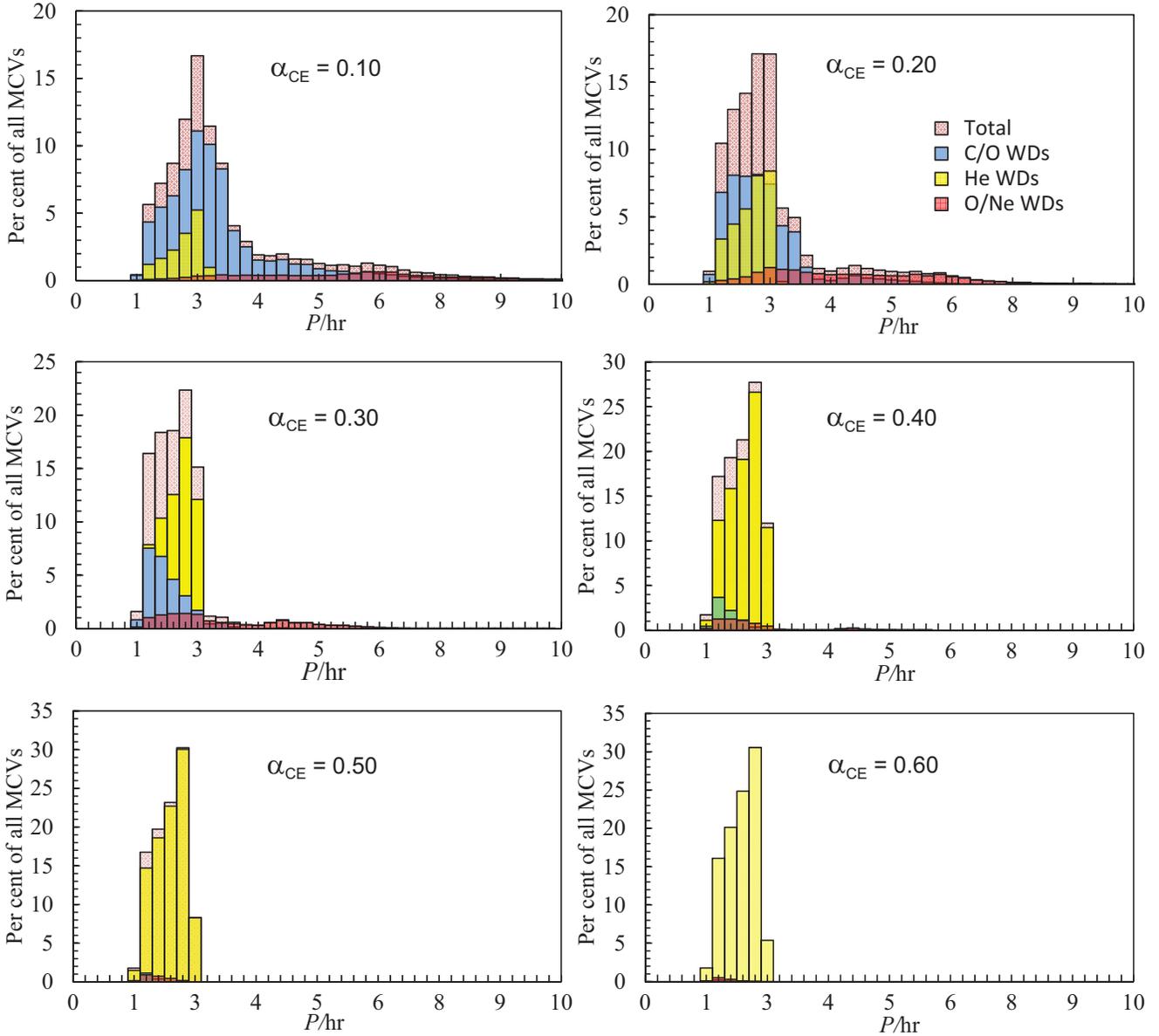}
\caption{Theoretical period distribution of magnetic systems just
  before they start RLOF for various $\alpha$'s.
  The period distribution of the primary white dwarf types is shown as the
  superimposed coloured categories. The total of the distribution is
  shown as the pink background histogram peaking around 2.8 
  to 3.0\,hrs. This is to be compared with the observed distribution
  for PREPs in Fig.\,\ref{preps_fig}}
\label{fig:RLOFPeriodP}
\end{center}
\end{figure*}

\begin{figure*}
\begin{center}
\includegraphics[width=1.00\textwidth]{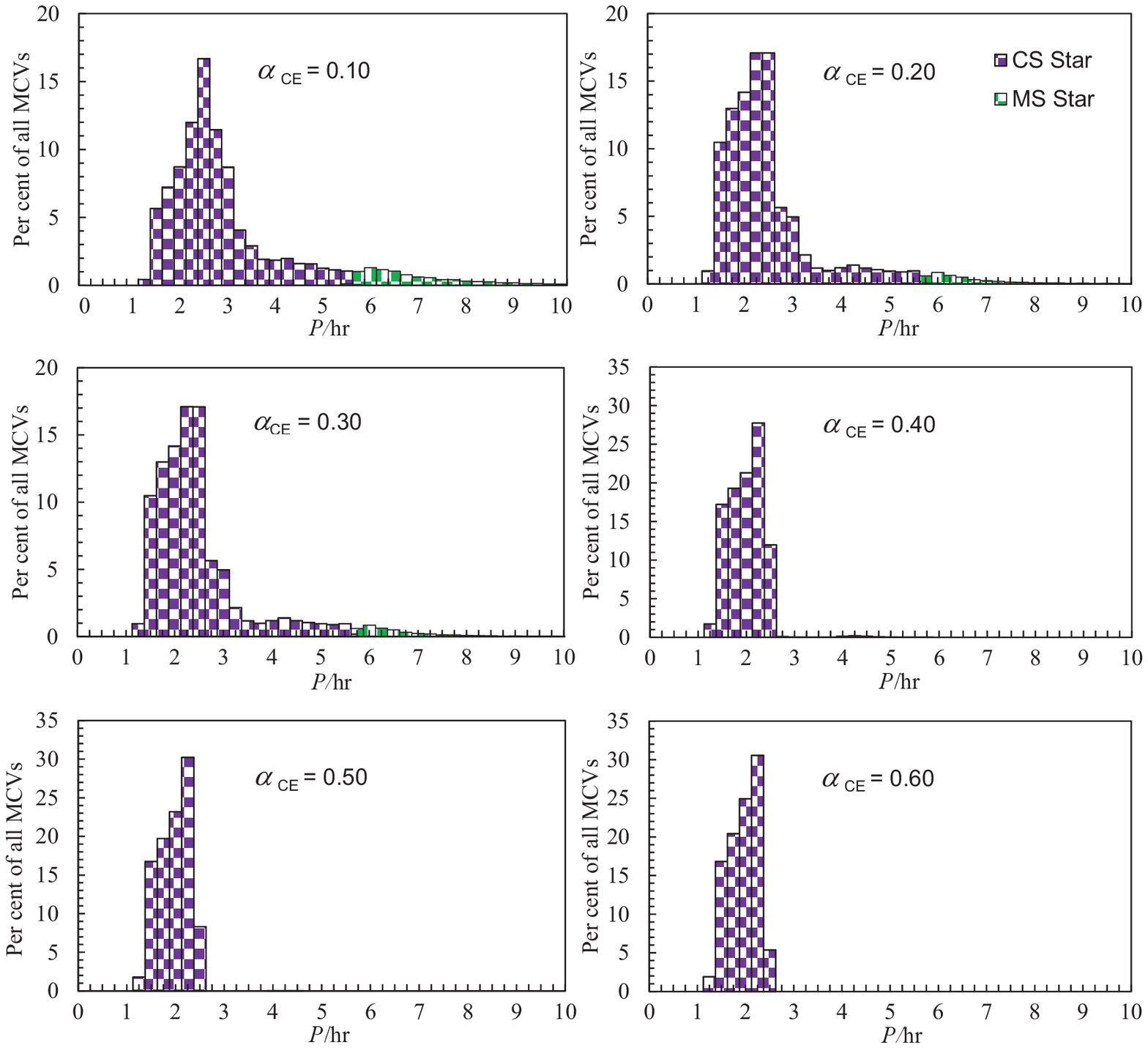}
\caption{Same as Fig.\ref{fig:RLOFPeriodP} but with the secondary star
  types shown as the superimposed coloured categories.  Both secondary
  star types are MS stars.  The CS type is a deeply or fully
  convective MS star with $M<0.7$\,M$_\odot$.}
\label{fig:RLOFPeriodS}
\end{center}
\end{figure*}

\subsubsection{Period distribution}\label{PerDist}

Figs\,\ref{fig:RLOFPeriodP} and \ref{fig:RLOFPeriodS} show the
theoretical period distribution of the PREPs just before the beginning
of Roche lobe overflow (RLOF) in a present day population formed over
the age of the Galactic Disc for various $\alpha$. The contributions
to the period distribution by white dwarf primaries of a certain
type are depicted in Fig.\,\ref{fig:RLOFPeriodP} while the
contributions to the period distribution by the secondaries of a given
type are displayed in Fig.\,\ref{fig:RLOFPeriodS}.

The period distribution peaks around $3\,$hr with a long tail
extending to about $10\,$hr for low $\alpha$. We note that at low
$\alpha$ our synthetic population tends to have orbital periods
clustering around the 1 to 4\,hr region while at higher $\alpha$ they
are confined to the 1 to 3\,hr region.

Fig.\,\ref{fig:RLOFPeriodP} shows that when $\alpha=0.1$ the main
contributors to the whole range of periods are systems with CO\,WD
primaries characterised by orbital periods from about 1 to 7\,hr and a
peak near 3\,hr. Systems with He\,WDs are also present but are fewer
and their periods are below 3\,hr.  Massive Oxygen-Neon white dwarf
(ONe\,WD) primaries form a much smaller fraction of the distribution,
as expected from Salpeter's initial mass function, but make some
contribution to the full range of periods when $\alpha< 0.4$.

As $\alpha$ increases the fraction of CO\,WD systems decreases until
these all but disappear for $\alpha>0.5$ while the percentage of
He\,WDs increases dramatically. For $\alpha\ge 0.4$, the orbital
periods are always below 3\,hr and He\,WD systems well and truly
dominate the period distribution.  For $\alpha>0.5$ the only systems
that are predicted to exist are those with He\,WDs. The fraction of
ONe\,WD systems reaches a maximum near $\alpha=0.2$ and then
decreases.

We note that systems with He\,WDs tend to populate the lowest period
range at all $\alpha$. These systems are generally characterised by
initially lower-mass primaries, and thus lower-mass companions
because $q\le 1$, and shorter orbital periods and initiate common
envelope evolution before helium ignition. Usually systems
characterised by short initial periods are unlikely to survive at low
$\alpha$ because the stronger drag force exerted on the two stars
during common envelope evolution causes them to merge.

Fig.\,\ref{fig:RLOFPeriodS} shows that most companions, particularly
at shorter orbital periods, are low-mass deeply convective stars.
More massive secondaries are generally found at longer periods for
three reasons. First, longer orbital periods require high-mass white
dwarfs to initiate stable mass transfer over the age of the Galactic
Disc and these massive white dwarfs can have secondaries with
masses all the way up to $1.44$\,\msun, provided $q\le 1$. Second,
during common envelope evolution for a fixed primary initial mass and
orbital period, systems with more massive secondaries have more
orbital energy and so a smaller portion of this energy is necessary to
eject the envelope. This leads to longer orbital periods. Third, for
a fixed white dwarf mass, more massive secondaries fill their Roche
lobes at longer orbital periods and so systems with more massive
companions are more likely to evolve into PREPS.

\subsubsection{Stellar pair distribution}\label{stellarpairs}

Table~\,\ref{tab:Contribs} lists fractions of the various combinations
of types of white dwarf primaries and secondary types just before RLOF
commences. At low $\alpha$ the predominant combination is a CO\,WD
primary with a low-mass M-dwarf secondary.  Second in abundance are
systems comprised of a He\,WD with a low-mass M-dwarf secondary.
Other combinations are also found but in much smaller numbers.  At
high $\alpha$ the two major categories are swapped and those systems
with He\,WD primaries become the predominant type. The observed
fraction of He\,WDs ($f_{\rm He}$) is generally low among classical
CVs ($f_{\rm He}\lsimeq 10$\,per cent) and pre-CVs
($f_{\rm He}\lsimeq 17\pm8$\,per\,cent as shown by
\citet{Zorotovic2011}. The results in Table\,\ref{tab:Contribs}
indicate that in order to reproduce the observed low fraction of
He\,WDs our models need to be restricted to $\alpha<0.3$.

\begin{table*}
  \caption{The fraction of the
    combinations of types of white dwarf primaries and
    secondary types just before RLOF commences for various $\alpha$.
    The stellar type CS is a deeply or fully convective low-mass MS star with $M<0.7$\,M$_\odot$.}
  \centering
 \label{tab:Contribs}
\begin{tabular}{c c c c c c c }
\hline
\noalign{\smallskip}
 \multicolumn{7}{c}{MCV progenitor pairs, fraction per cent} \\
$\alpha$  &  He\,WD/CS & CO\,WD/CS & ONe\,WD/CS & He\,WD/MS & CO\,WD/MS & ONe\,WD/MS \\
\noalign{\smallskip}
\hline
\noalign{\smallskip}
0.10  &14.86  &69.63  & 5.72   &0.00  &3.77  &6.03 \\
0.20  &30.27  &52.27  &12.99  &0.00  &0.38  &4.10 \\
0.30  &61.36  &25.69  &12.49  &0.00  &0.00  &0.46 \\ 
0.40  &96.44  & 7.78  & 5.78   &0.00  &0.00  &0.00 \\ 
0.50  &95.85  & 1.72  & 2.44   &0.00  &0.00  &0.00 \\ 
0.60  &98.75  & 0.28  & 0.98   &0.00  &0.00  &0.00 \\ 
0.70  &99.67  & 0.01  & 0.32   &0.00  &0.00  &0.00 \\ 
0.80  &99.92  & 0.00  & 0.00   &0.00  &0.00  &0.00 \\ 
\hline
\end{tabular}
\end{table*}

\subsubsection{Mass distribution}\label{mass_distribution}

\begin{figure*}
\begin{center}
\includegraphics[width=1.01\textwidth]{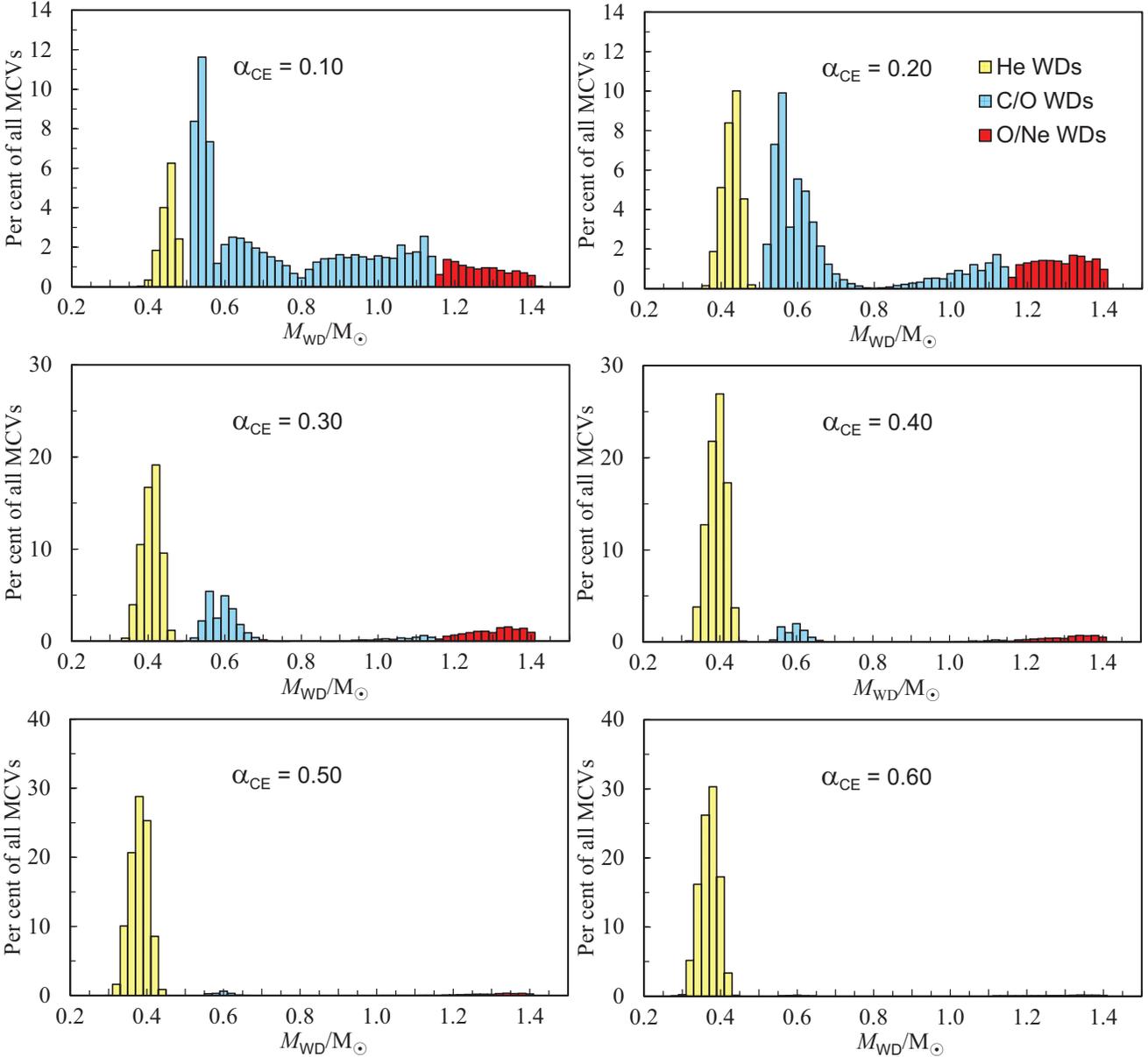}
\caption{Theoretical mass distribution of the white dwarf primary star
  of magnetic systems just before they start RLOF for various
  $\alpha$. The distributions of the three white dwarf types are shown
  as three superimposed coloured categories.}
\label{fig:RLOFMassP}
\end{center}
\end{figure*}

Fig.\,{\ref{fig:RLOFMassP}} shows that all our models predict that, on
average, longer orbital period systems contain CO\,WDs while
shorter-period systems tend to have He\,WD primaries. At low $\alpha$
the primaries are predominantly CO\,WDs with masses in the range 0.5
to 1.1\,M$_\odot$ followed in lesser numbers by He\,WDs with masses in
the range 0.4 to 0.5\,M$_\odot$ while ONe\,WDs, with masses in the
range 1.2 to 1.4\,M$_\odot$, are rarer with their incidence reaching a
maximum near $\alpha=0.2$.

We note that there is a curious dip in the white dwarf mass
distribution near $M_{\rm WD}=0.8$\,M$_\odot$ which widens as
$\alpha$ increases until all CO and ONe\,WDs disappear for
$\alpha> 0.5$.  This is because as $\alpha$ increases,
systems emerge from common envelope at progressively longer periods,
because large $\alpha$ means a high envelope clearance efficiency
which leads to larger stellar separation at the end of the common
envelope stage. However the longer the orbital period, the higher the
white dwarf mass needs to be for stable mass transfer to
commence. Thus the gap in the white dwarf mass distribution is caused
by those systems that emerge from common envelope at large separations
but with white dwarf primaries that are not massive enough to allow
RLOF to take place. Another, albeit much narrower, gap occurs near
0.5\,M$_\odot$ for all $\alpha$ but becomes wider for
$\alpha\ge0.2$. This gap also persists until all CO and ONe\,WDs
disappear at $\alpha > 0.5$. It divides systems with He\,WDs primaries
from those with CO\,WDs and is linked to whether the stars enter
common envelope evolution on the RGB, and so produce a He\,WD primary
with $M_{\rm WD}\lsimeq 0.5$\,M$_\odot$, or on the AGB, and so produce
a CO\,WD primary with $M_{\rm WD}> 0.5$\,M$_\odot$.

\begin{figure*}
\begin{center}
\includegraphics[width=1.00\textwidth]{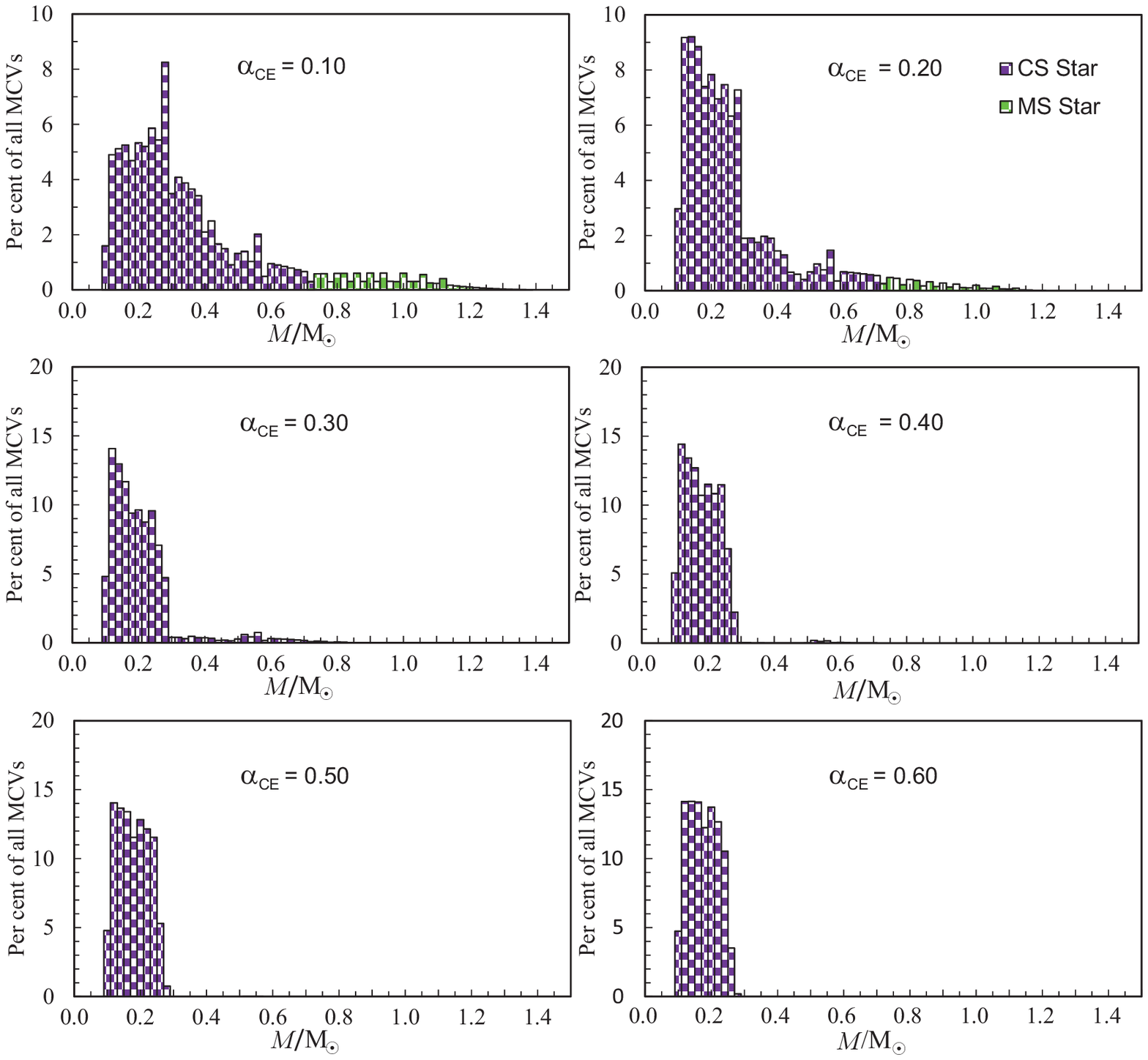}
\caption{As in Fig\,\ref{fig:RLOFMassP} but for the secondary star
  types shown as the coloured categories. Both secondary star types
  are MS stars.  The CS type is a deeply or fully
  convective MS star with $M<0.7$\,M$_\odot$.}
\label{fig:RLOFMassS}
\end{center}
\end{figure*}

Fig.\,\ref{fig:RLOFMassS} shows again that the secondaries are
predominantly low-mass deeply or fully convective M-dwarf stars. The
distribution has a broad peak around 0.1 to 0.3\,M$_\odot$ at
$\alpha=0.1$ to 0.2 with a long tail extending to $1.2$\,M$_\odot$. As
$\alpha$ increases, the peak in the secondary mass distribution shifts
to slightly lower masses (around 0.1 to 0.25\,M$_\odot$) but the
high-mass tail shrinks quite dramatically. At $\alpha\ge0.4$ the
distribution is confined to secondary masses of less than about
$0.3$\,M$_\odot$. As already noted in section \ref{PerDist}, the
majority of these very low-mass donor stars belong to systems that
underwent common envelope evolution during the Hertzsprung gap or RGB
phases and thus have He\,WD primaries with
$M_{\rm WD}\lsimeq 0.5$\,M$_\odot$.  We also note that systems with
low-mass secondaries ($M_{\rm sec}\lsimeq 0.35$\,\msun) remain detached
for longer because magnetic braking is inefficient in these stars and
gravitational radiation is the main source of loss of angular
momentum.

\subsubsection{Magnetic field distribution}\label{MagFieldDistr}

Fig. \ref{fig:RLOFOvTBmag} shows the theoretical magnetic field
distribution and the breakdown of the primary white dwarf types for
our range of $\alpha$.  The maximum field strength is a few $10^8$\,G
and is found mostly in systems whose primary is a He\,WD. The reason
for this is that systems that undergo common envelope evolution during
the RGB evolution have shorter initial orbital periods and create very
short period binaries with a highly magnetic white dwarf, as expected
from equation (\ref{EqBfield}).

The magnetic field distribution is dominated by systems with CO\,WD
primaries when $\alpha\le 0.2$. When $\alpha\ge 0.4$ the field
distribution becomes narrower and its peak shifts to higher field
strengths. For $\alpha\ge0.5$ the field distribution only contains
very highly magnetic He\,WD primaries with a peak near
$3.2\times 10^8$\,G. This shift to high fields is because those
systems that go through common envelope evolution while their
primaries are on the RGB merge for low $\alpha$ but can survive for
high $\alpha$ giving rise to very short orbital period systems with
strongly magnetic, low-mass white dwarfs.

We note that the magnetic field distribution has a dip near
$8\times 10^6$\,G appearing at $\alpha\ge 0.2$ and persisting until
all CO and ONe\,WDs disappear from the distribution. This is
reminiscent of the dip that we encountered in the white dwarf mass
distribution (see \ref{mass_distribution}) and has the same
explanation. The similar behaviour is because the magnetic field
strength is a function of white dwarf mass (by virtue of equations
\ref{EqBfield}). The field dip is thus
caused by the dearth of systems with white dwarf masses around
$0.8\,\msun$ (see Fig.\,\ref{fig:RLOFMassP}).

\section{Comparison to observations}\label{Comparison}

The optimal observational sample with which to compare our results
would be that formed by the known magnetic PREPs. However, this sample
is exceedingly small and observationally biased. To make things worse,
not all PREPS have well determined parameters, such as masses and
magnetic field strengths. Hence, for some of these studies we use the
observed sample of MCVs, noting the following important points (i) the
MCV sample is magnitude-limited, (ii) MCVs suffer from prolonged high
and low states of accretion and (iii) MCVs include systems at all
phases of evolution. Some of them began Roche lobe overflow billions
of years ago while others have only recently begun mass
exchange. Therefore, one should take such a comparison with some
degree of caution particularly when we compare quantities
that change over time, such as orbital periods and masses. When
comparing masses we will also use the observed sample of non-magnetic
Pre-CVs \citep{Zorotovic2011}.

\begin{figure*}
\begin{center}
\includegraphics[width=1.0\textwidth]{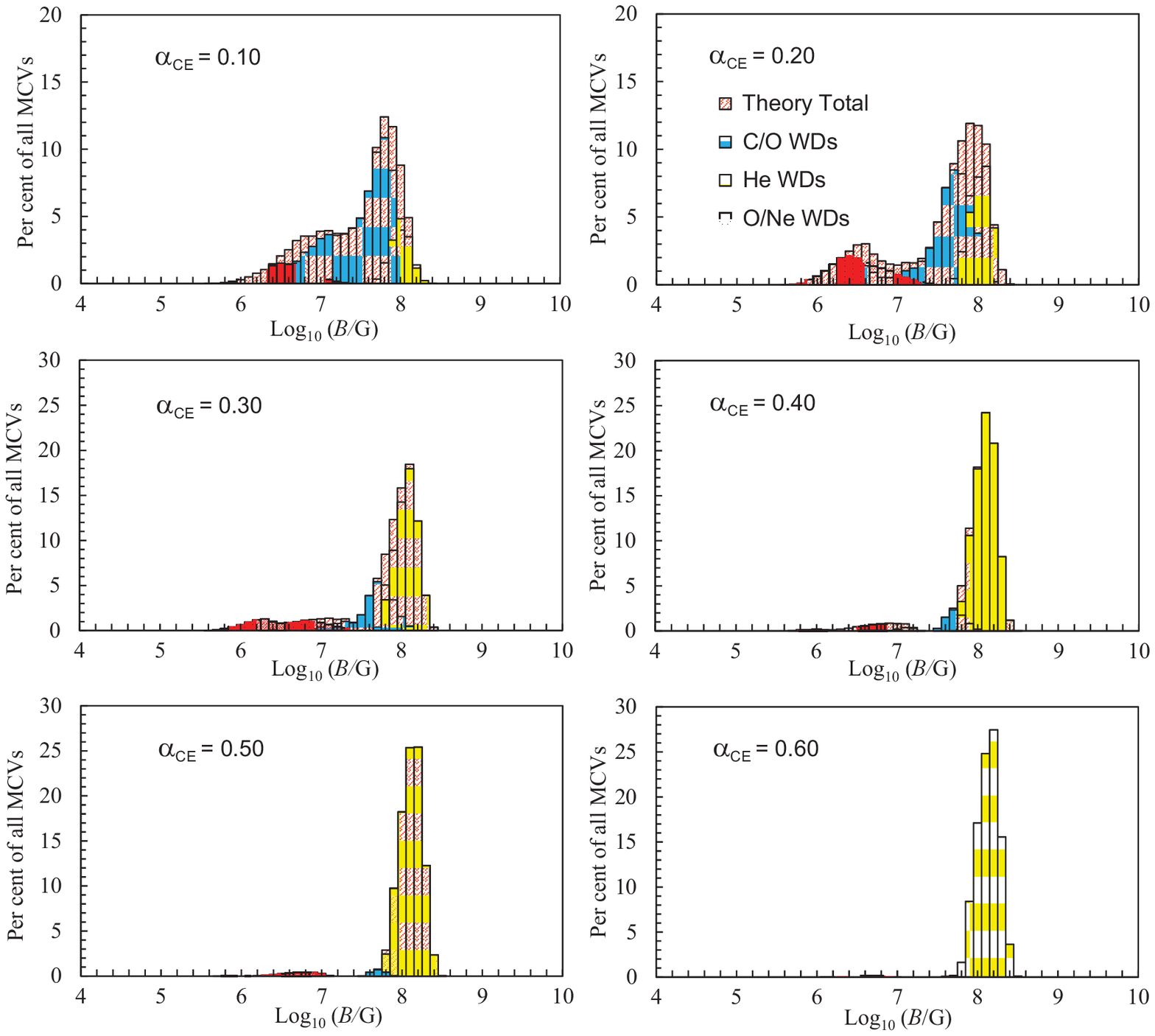}
\caption{Pink shaded histogram: Total theoretical magnetic field
  distribution of the white dwarf primary stars in magnetic systems just before they
  start RLOF for the indicated $\alpha$. The histograms of the three
  types of white dwarfs making up the total theoretical magnetic field distribution are
  shown as the foreground coloured histograms.  These three are made partially
  transparent so that details of the other histograms can be seen through them.}
\label{fig:RLOFOvTBmag}
\end{center}
\end{figure*}

The tables of \citet{fer2015a} show that the observed orbital periods
of MCVs are in the range 1 to 10\,hr, masses are between about $0.4$
and $1.1$\,$\msun$ and that the magnetic field distribution is relatively
broad with a peak near $3.2\times10^7$\,G.  A quick glance at
Figs\,\ref{fig:RLOFPeriodP}, \ref{fig:RLOFMassP} and
\ref{fig:RLOFOvTBmag} immediately reveals that models with
$\alpha>0.3$ are all unable to reproduce the general
characteristics expected from the progenitors of the observed
population of MCVs and we elaborate on this in more details
below. Generally, we see that models with $\alpha>0.3$ are
not realistic and evolutionary effects cannot account for the large
degree of discrepancy between theory and observations.

We begin our analysis with the magnetic field distribution. There is
no evidence for field decay among MCVs \citep{fer2015a,Zhang2009} so we can
assume that the magnetic field strength remains unchanged over the
entire life of the magnetic binary.

\begin{figure}
\begin{center}
\includegraphics[width=1.00\columnwidth]{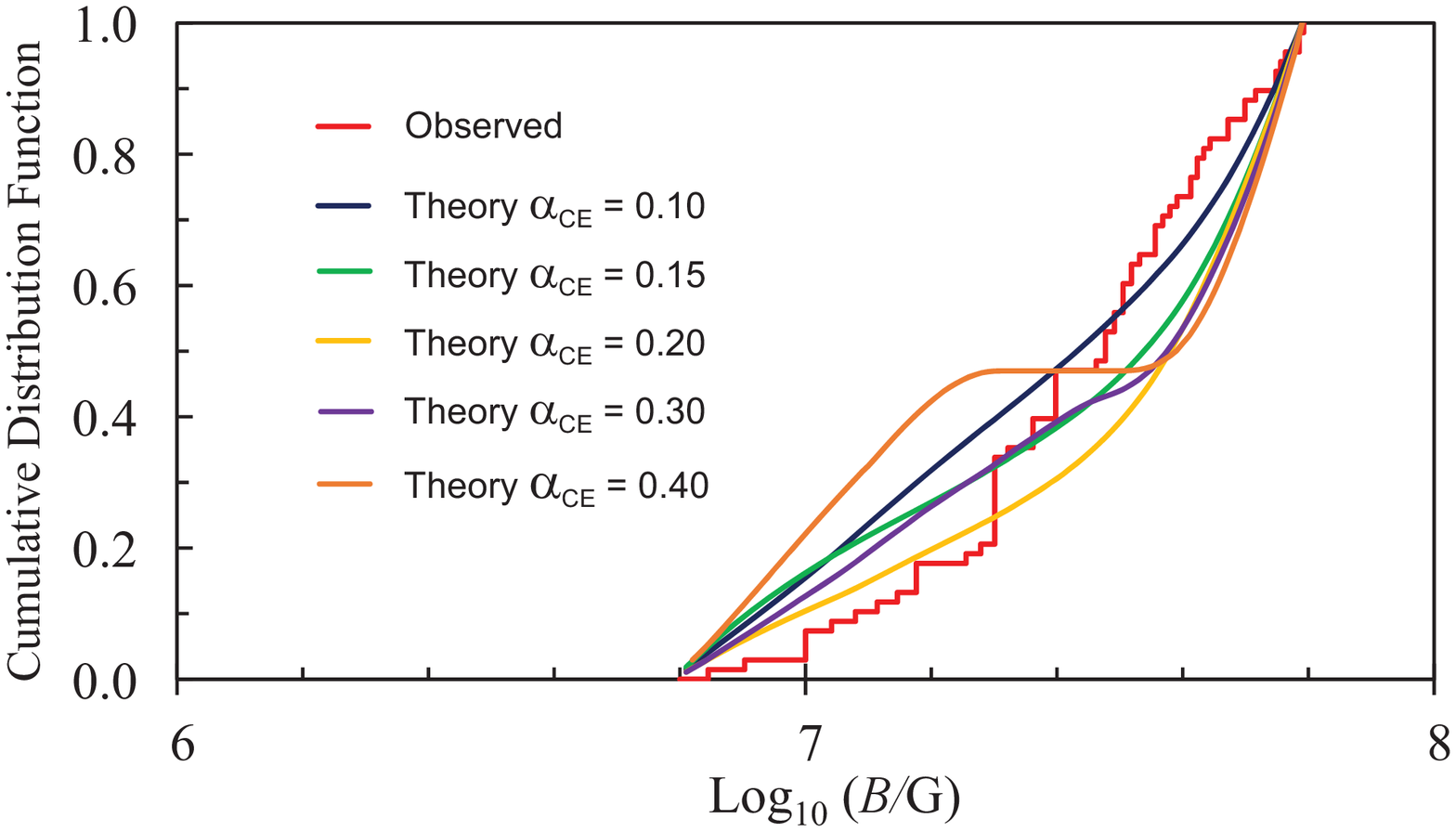}
\caption{Theoretical cumulative distribution functions for the
  magnetic fields of MCV white dwarfs at RLOF for 
  $\alpha=0.10, 0.15, 0.20, 0.30$ and $0.40$ and the CDF of the
observed magnetic field of 81 systems taken from \citet{fer2015a} }
\label{fig:MCVWDBMagCDF}
\end{center}
\end{figure}

We have used a Kolmogorov--Smirnov (K--S) test \citep{Press1992} to
compare the magnetic field distribution of the observed population
with the theoretical results. This test establishes the likelihood
that two samples are drawn from the same population by comparison of
the cumulative distribution functions (CDFs) of the two data
samples. The CDFs of the two distributions vary between zero and one
and the test is on the maximum of the absolute difference $D$ between
the two CDFs. It gives the probability $P$ that a random selection
would produce a larger $D$. Five model CDFs for five different
$\alpha$'s and the CDF for the known observed magnetic fields of 81 MCV
systems are compared in Fig.\,\ref{fig:MCVWDBMagCDF}.

\begin{table}

\caption{Kolmogorov–-Smirnov $D$ statistic and probability $P$ of the magnetic field
distributions of the  observed and synthetic populations of MCVs for a range of $\alpha$.}
\centering
\begin{tabular} { c c c}
\hline
\noalign{\smallskip}
$\alpha$ & $D$  & $P$ \\
\noalign{\smallskip}
\hline
\noalign{\smallskip}
0.10 & 0.17476 & 0.36069 \\
0.15 & 0.19349 & 0.24632 \\
0.20 & 0.25141 & 0.05845 \\
0.30 & 0.22962 & 0.10500 \\
0.40 & 0.26939 & 0.04298 \\
0.50 & 0.35186 & 0.00429 \\
0.60 & 0.38035 & 0.00006 \\
0.70 & 0.61987 & 0.00000 \\ 
0.80 & 0.94366 & 0.00000 \\
\noalign{\smallskip}
\hline
\end{tabular}
\label{tab:KS_Bfield}
\end{table}

The observed samples of MCVs and magnetic PREPs are very biased,
particularly at the low and high ends of the magnetic field
distribution. At low fields ($B\lsimeq 10$\,MG) the observed radiation
is dominated by the truncated accretion disc. In these low-field
systems the photosphere of the white dwarf is never visible and Zeeman
splitting cannot be used to determine field strengths. Nor can
cyclotron lines be used to measure fields because they are too weak
and invisible in the observed spectra. In the high field regime
($B\gsimeq 100$\,MG) mass accretion from the companion star is
inhibited \citep{Ferrario1989, LiWuWick1994} and so high field MCVs
are very dim wind accretors often below the detection limits of most
surveys \citep[AR\,UMa][]{Hoard2004}. Because of these biases the
observed samples in these regimes are far from complete and
theoretical fits are unreliable. We therefore restrict our comparison
between theory and observations to field strengths in the range 10 to
70\,MG.

\begin{figure}
\begin{center}
\includegraphics[width=1.0\columnwidth]{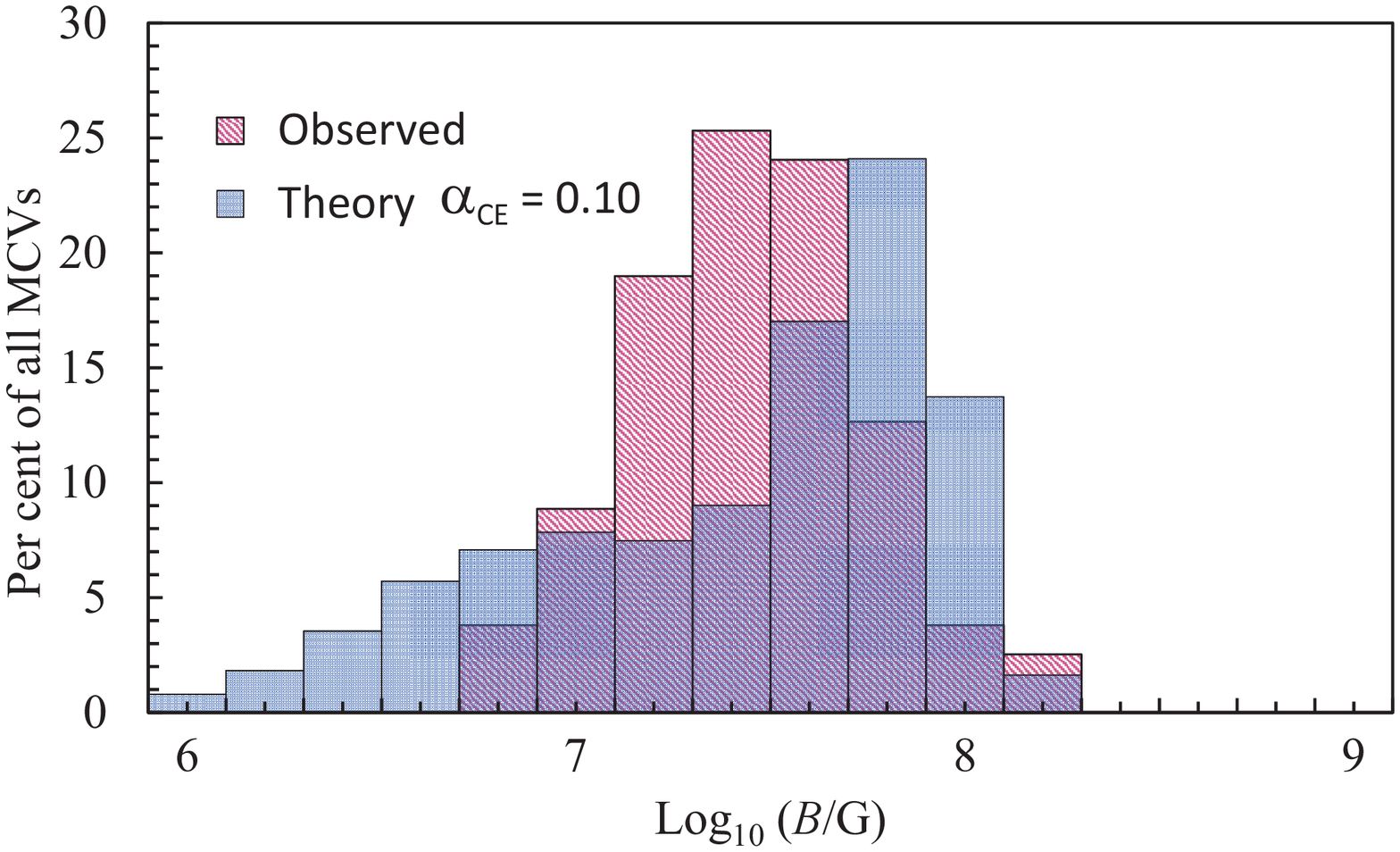}
\caption{Comparison of the theoretical magnetic field strength for
  $\alpha= 0.1$ and the observed magnetic field strength of the 81
  MCVs taken from \citet{fer2015a}}
\label{fig:CompareBmag}
\end{center}
\end{figure}

The results of the K--S test for our range of $\alpha$ are displayed
in Table \ref{tab:KS_Bfield} and show that the field distribution is a
better match to the observations at low $\alpha$. The comparison of
the magnetic field distribution between theory and observations is
shown in Fig.\,\ref{fig:CompareBmag} for $\alpha=0.1$.

We stress that for $\alpha>0.3$ all the theoretical magnetic field
distributions shown in Fig.\,\ref{fig:RLOFOvTBmag} are very
unrealistic because only very high field ($B>60$\,MG) He\,WDs
($M\lsimeq0.5$\,\Msun) are predicted to exist.  This is contrary to
observations that show that fields cover a much wider range of
strengths ($\mbox{a few}10^6$ to a few $\,10^8$\,G) and white dwarf
masses (0.4 to 1.2\,\Msun) as seen in Tables\,2 and 3 of
\citet{fer2015a}. We note here that a larger $B_0$ with
  $\alpha>0.3$ is not a good fit because high $\alpha$ models have a
  very narrow distribution of field strengths skewed to high fields. A
  larger $B_0$ would only exacerbate this by pushing the distribution
  to even higher fields. If $B_0$ were chosen at the lower end of the
  acceptable range found in paper\,II the field distribution would
  shift to lower fields and yield an average field strength closer to
  observations. However, because $B_0$ can only shift the field
  distribution to lower (or higher) fields, the very narrow width of
  the distribution that characterises all our high $\alpha$ models
  would not be corrected by readjusting $B_0$. We can therefore
  conclude that the $B_0$ calibration carried out in paper\,II holds
  for the modelling of magnetic white dwarfs in binaries and supports
  our common envelope dynamo theory, according to which the fields of
  MCVs are scaled down, through equation\,(\ref{EqBfield}),
  from the maxima obtained during merging events that generate
  isolated HFMWDs.

We have performed a K--S study between the synthetic white dwarf mass
distribution and that of white dwarf masses in MCVs taken from
\citet{fer2015a}. In principle such a comparison can be
justified if we make the usual assumption that the mass of the white
dwarf does not grow in CVs because nova eruptions tend to expel all
material that is accreted. However, we have found that the
K--S test applied to the white dwarf masses of the theoretical and
observed population of MCVs yields poor results, as shown in the
second and third columns of Table \ref{tab:KS_WDmass}. Such
a conflict is not surprising because our assumption that the mass of
the white dwarf does not grow because of nova eruptions may not be
correct.

In this context, we note that \citet{Zorotovic2011} noticed a curious
discrepancy in their observational data of CVs and Pre-CVs. That is,
they found that the mean white dwarf mass in CVs
($0.83\pm 0.23$\,M$_\odot$) significantly exceeds that of pre-CVs
($0.67\pm0.21$\,\,M$_\odot$) and they excluded that this difference
could be caused by selection effects.  The two possible solutions
advanced by \citet{Zorotovic2011} were that either the mass of the
white dwarf increases during CV evolution, or a short phase of thermal
time-scale mass transfer comes before the formation of CVs during
which the white dwarf acquires a substantial amount of mass via stable
hydrogen burning on the surface of the white dwarf \citep[as first
suggested by][]{Schenker2002}. During this phase the
system may appear as a super-soft X-ray source
\citep[][]{Kahabka1997}. Using this assumption \citet{Wijnen2015}
could build a large number of massive white dwarfs. However their
model still created too many low-mass He\,WDs and too many evolved
companion stars contrary to observations. Another possibility has
recently been advanced by \citet{Zorotovic2017}.  In order to achieve
a better agreement between their binary population synthesis models
and observations of CVs they adopted the ad-hoc mechanism proposed by
\citet{Schreiber2016} which surmises the existence of additional
angular momentum losses generated by mass transfer during the CV
phase. Such losses are assumed to increase with decreasing white dwarf
mass and would cause CVs with low-mass white dwarfs to merge and
create an isolated white dwarf. By removing these merged systems from
the synthetic CV sample the average white dwarf mass
increases. Furthermore such a mechanism would explain the existence of
isolated low-mass white dwarfs ($M_{\rm WD}<0.5\,\msun$) that
constitute around 10\,per cent of all single white dwarfs observed in
the solar neighbourhood \citep[e.g.][]{Kepler2007}.

\begin{figure}
\begin{center}
\includegraphics[width=1.0\columnwidth]{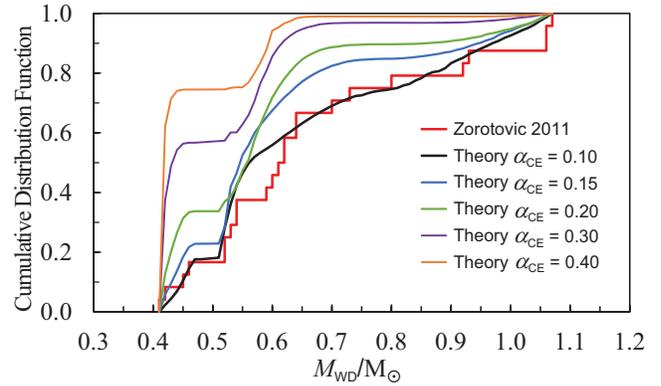}
\caption{Cumulative Distribution Functions of the mass distributions
  for the observed pre-CV white dwarf masses taken from
  \citet{Zorotovic2011} and the theoretical distribution of the
  white dwarfs as the systems start RLOF for $\alpha=0.10,
  0.15, 0.20, 0.30$ and $0.40$.  The K-S statistics for this plot are
  shown in the fourth and fifth columns of table~\ref{tab:KS_WDmass}}
\label{fig:MCV-WD-Mass-CDF}
\end{center}
\end{figure}

\begin{figure}
\begin{center}
\includegraphics[width=1.0\columnwidth]{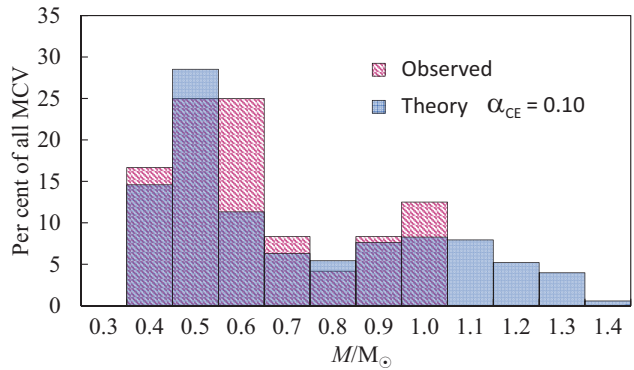}
\caption{Comparison of the mass distributions for the observed pre-CV white
dwarf masses taken from \citet{Zorotovic2011} and the theoretical mass
distribution of the white dwarfs as the systems start RLOF for
$\alpha= 0.10$.}
\label{fig:CompareMasses}
\end{center}
\end{figure}

Going back to our studies, if a comparison between white dwarf masses
in MCVs and our synthetic population may not be meaningful, the next
best sample to use for our K--S test is the observed white dwarf
masses of pre-CVs \citep{Zorotovic2011}.  We show in
Fig.\,\ref{fig:MCV-WD-Mass-CDF} the CDFs of the mass distributions
for the observed sample and the theoretical distribution of the white
dwarf masses at the beginning of RLOF for various $\alpha$. The results are
reported in the fourth and fifth columns of Table \ref{tab:KS_WDmass}
and show that the agreement between observations and theory is greatly
improved.  The comparison of the synthetic and observed Pre-CV\, white
dwarf mass distribution is shown in Fig.~\ref{fig:CompareMasses} for
the largest K--S probability at $\alpha = 0.10$. This suggests that
the white dwarf mass distribution of the binary population in which
primaries develop a magnetic field during common envelope evolution
and later evolve to MCVs does not differ from that of the binary
precursors of classical non-magnetic CVs.

\begin{table}
  \caption{K--S $D$ statistic and probability $P$ of the
    white dwarf mass distributions of the observed MCVs listed by
    Ferrario et al. (2015a, second and third
    columns) and our synthetic populations
    for $\alpha$ given in the first column. In the fourth and fifth
    columns we show the K--S results of the observed
    Pre-CV masses of \citet{Zorotovic2011} and our synthetic
    populations at the start of RLOF (fourth and fifth columns).}
  \centering
\begin{tabular} {c c c c c}
\hline
\noalign{\smallskip}
$\alpha$&$D$&$P$&$D$&$P$\\
\noalign{\smallskip}
\hline
\noalign{\smallskip}
0.10 & 0.37687  & 0.02023088 & 0.12954  & 0.95281557 \\
0.15 & 0.49861  & 0.00064407 & 0.23478  & 0.34844783 \\
0.20 & 0.56677  & 0.00006150 & 0.26010  & 0.23507547 \\
0.30 & 0.62615  & 0.00000622 & 0.48014  & 0.18713800 \\ 
0.40 & 0.69590  & 0.00000032 & 0.66148  & 0.00106500 \\ 
\noalign{\smallskip}
\hline
\end{tabular}
\label{tab:KS_WDmass}
\end{table}

We note that the Pre-CV observational sample shows a dearth of systems
in the white dwarf mass distribution centred around $0.8\,\msun$. This
mass gap was already noted in the theoretical {\sc{bse}} models and the
reasons for its existence were explained in section \ref{mass_distribution}.
The smaller size of this gap for models with $\alpha\le0.2$ explains why
we achieve a better fit with observations for $\alpha=0.1$, as indicated
by the K--S test.

If $\alpha>0.3$ the theoretical white dwarf mass distribution shown in
Fig.\,\ref{fig:RLOFMassP} is very unrealistic because only He\,WDs
($M<0.5$\,\Msun) are predicted to exist by these models.  This is
contrary to observations that show that masses cover the much wider
range 0.4 to 1.2\Msun \citep[see Tables\,2 and 3 in][]{fer2015a}.

Next, we look at the secondary mass distribution, keeping in mind that
a comparison between our synthetic {\sc{bse}} mass sample and the
observed secondary masses in MCVs is definitely not appropriate
because secondary masses decrease over time as mass is transferred to
the white dwarf. Nonetheless it may still be pertinent to use the
observed Pre-CV sample to study and compare the overall
characteristics of these samples so that we can, at the very least,
discard some of the most extreme theoretical models.

Fig. \ref{fig:RLOFMassS} shows that if $\alpha>0.3$ then
$M_{\rm sec}< 0.3\,\msun$, which is inconsistent with observations of
pre-CVs \citep[see][]{Zorotovic2011}. Furthermore, we can see that
when $\alpha>0.2$, the decline towards higher masses becomes far too
steep. This straightforward comparison seems again to indicate that
models with $\alpha>0.3$ are very unrealistic and therefore low $\alpha$
yields a better fit to observations.

\section{Discussion and Conclusions}\label{DandC}

The origin of large-scale magnetic fields in stars is still a puzzling
question \citep[see][]{fer2015b}. However, the results from
recent surveys such as the SDSS \citep{Kepler2013}, BinaMIcS
\citep{ale2015} and MiMes \citep{Wade2016} have provided us with a much
enlarged sample of magnetic stars that have allowed investigators to
conduct studies like this one. There are two main competing scenarios
to explain the existence of magnetic fields in white dwarfs. In 1981,
\citeauthor{ang1981} proposed that the magnetic Ap and Bp stars
are the most likely progenitors of the highly MWDs under the
assumption of magnetic flux conservation \citep[see
also][]{tout2004,wickramasinghe2005}. 

The best clue so far on the origin of fields in white dwarfs (isolated
and in binaries) has come from the study of their binary properties
\citep{Liebert2005,Liebert2015}, as outlined in section
\ref{WhereProg}. This is why the proposal by \citet{Tout2008}, that
the origin of magnetic fields in white dwarfs is related to their
duplicity and stellar interaction during common envelope evolution, is
becoming more and more appealing.

We have extended our population synthesis study of binary systems
carried out for papers\,I and\,II for the HFMWDs to explain the origin
of fields in the accreting white dwarfs in MCVs.  Similarly to the
investigations conducted in paper\,I and\,II, we have varied the
common envelope efficiency parameter $\alpha$ to investigate its
effects on the resulting synthetic population of MCVs.  We have shown
that models with $\alpha\ge 0.4$ are not able to reproduce the large
range of white dwarf masses, field strengths, and secondary types and
masses that are observed in MCVs and therefore models with
$\alpha< 0.4$ best represent the observed data.  K--S tests conducted
to compare our synthetic white dwarf mass and magnetic field
distributions with the observed populations have given us some
quantitative support in favour of models with $\alpha<0.4$.

However, we need to stress again some of the shortcomings of our work
and in particular those that arise from our comparison to
observations. Many of the parameters (e.g., white dwarf mass, magnetic
field, secondary star mass and type, orbital period) that characterise
the Galactic populations of MCVs and PREPs and are needed for
comparison studies are often hard to ascertain owing to evolutionary
effects and observational biases that are difficult to
disentangle. For instance, we mentioned in section \ref{Comparison}
that magnetic white dwarfs in PREPs would be the best objects with
which to compare our theoretical results and in particular the mass
distribution, because mass is not contaminated by accretion
processes. On the other hand there are far too few members of this
population. The white dwarf mass distribution provided by the much
larger sample of MCVs cannot be used either for comparison purposes
because masses vary over time, owing to accretion and nova
explosions. So we have used the sample provided by the
non-magnetic Pre-CVs of \citet{Zorotovic2011}.

The situation is somewhat ameliorated when we consider the magnetic
field distribution because fields are not expected to change over time
\citep[see][]{fer2015a}. However, the true magnetic field distribution
of MCVs is not well known because it is plagued by observational
biases. For example, at field strengths below a few\,$10^7$\,G most
systems (the intermediate polars) have an accretion disc from which
continuum emission and broad emission lines swamp the Zeeman and
cyclotron features arising from the white dwarf surface
and so hide those spectral signatures that are
crucial to determine their field strengths. Very high field polars are
also likely to be under-represented in the observational set because
mass accretion from the companion star is impeded by the presence of
strong fields \citep{Ferrario1989, Li1998} making these systems
very dim wind accretors.

Despite the limitations highlighted above, we have shown that the
characteristics of the MCVs are generally consistent with those of a
population of binaries that is born already in contact (exchanging
mass) or close to contact, as first proposed by \citet{Tout2008}. This
finding is also in general agreement with the hypothesis that the
binaries known as PREPs, where a MWD accretes matter from the wind of
a low-mass companion, are the progenitors of the MCVs.

\section*{Acknowledgements}

GPB gratefully acknowledges receipt of an Australian Postgraduate
Award.  CAT thanks the Australian National University for supporting a
visit as a Research Visitor of its Mathematical Sciences Institute,
Monash University for support as a Kevin Westfold distinguished visitor
and Churchill College for his fellowship.\\

\end{document}